\documentclass[a4paper]{jpconf}
\usepackage{graphicx}
\usepackage{amsmath}
\usepackage{xcolor}
\begin{document}
\newcommand{\TeV}{\ensuremath{{\rm TeV}}}
\newcommand{\GeV}{\ensuremath{{\rm GeV}}}
\newcommand{\vtb}{\ensuremath{V_{\rm tb}}}
\newcommand{\twb}{\ensuremath{{\rm tWb}}}
\newcommand{\mandelt}{\ensuremath{t}}
\newcommand{\Wb}{\ensuremath{\rm W}}
\newcommand{\mandels}{\ensuremath{s}}
\newcommand{\tq}{\ensuremath{\rm t}}
\newcommand{\Pt}{\ensuremath{p_{\rm T}}}
\newcommand{\bq}{\ensuremath{\rm b}}
\newcommand{\mytt}{\ensuremath{\rm t\bar{t}}}
\newcommand{\invf}{\ensuremath{{\rm fb^{-1}} }}
\newcommand{\ETslash}{\ensuremath{\not\!\!E_{\rm T}}}
\newcommand{\vl}{\ensuremath{f_{\rm V}^{\rm L}}}
\newcommand{\vr}{\ensuremath{f_{\rm V}^{\rm R}}}
\newcommand{\gl}{\ensuremath{f_{\rm T}^{\rm L}}}
\newcommand{\gr}{\ensuremath{f_{\rm T}^{\rm R}}}
\newcommand{\cosTheta}{\ensuremath{\cos{\theta^*_\ell}}}
\newcommand{\fz}{\ensuremath{F_{\rm 0}}}
\newcommand{\myfl}{\ensuremath{F_{\rm L}}}
\newcommand{\fr}{\ensuremath{F_{\rm R}}}
\newcommand{\fzc}{\ensuremath{0.720}}
\newcommand{\fzcstat}{\ensuremath{0.039}}
\newcommand{\fzcsyst}{\ensuremath{0.037}}

\newcommand{\flc}{\ensuremath{0.298}}
\newcommand{\flcstat}{\ensuremath{0.028}}
\newcommand{\flcsyst}{\ensuremath{0.032}}

\newcommand{\frc}{\ensuremath{-0.018}}
\newcommand{\frcstat}{\ensuremath{0.019}}
\newcommand{\frcsyst}{\ensuremath{0.011}}

\newcommand{\cresfz}{\ensuremath{\fz=\fzc\pm\fzcstat\,{\rm (stat.)}\pm\fzcsyst\,{\rm (syst.)}}}
\newcommand{\cresfl}{\ensuremath{\myfl=\flc\pm\flcstat\,{\rm (stat.)}\pm\flcsyst\,{\rm (syst.)}}}
\newcommand{\cresfr}{\ensuremath{\fr=\frc\pm\frcstat\,{\rm (stat.)}\pm\frcsyst\,{\rm (syst.)}}}
\title{Single top measurements and the $|\vtb|$ extraction at the LHC}

\author{Abideh Jafari for the CMS and ATLAS collaborations}

\address{Universit\'e catholique de Louvain, Louvain-la-Neuve, Belgium}

\ead{abideh.jafari@cern.ch}

\begin{abstract}
The CMS and ATLAS experiments have performed detailed studies on the electroweakly produced top quarks at the LHC. These studies range from accurate measurements of the cross section and $|\vtb|$ in different production modes to search for new interactions in the $\twb$ vertex. Moreover, different properties of the top quark are precisely measured in this context. All measurements are consistent with the standard model and no sign of new physics is observed.
 
\end{abstract}

\section{Introduction}
At the LHC~\cite{JINSTLHC}, the top quark is mainly produced in pairs via the strong interaction. To lesser extent, it is produced individually through the electroweak interaction including the $\twb$. The $\mandelt$-channel process is the dominant single top production mode.
The $\Wb$-associated production (tW), recently observed at the LHC, occurs with a moderate rate whereas the $\mandels$-channel process is so rare that it has been only possible to set an upper limit for its cross section. In this article we report part of the latest LHC results on single top quark analyses. The analyses are performed using the proton-proton collisions recorded by the CMS~\cite{JINSTCMS} and ATLAS~\cite{JINSTATLAS} experiments at 7 and 8\,$\TeV$ center-of-mass energies.
\section{The $\mandelt$-channel cross section measurements}
The event signature of the $\mandelt$-channel single top in its leptonic final state contains a charged lepton (e or $\mu$), missing transverse energy ($\ETslash$) and two jets where one is originating from a b quark (b-tagged jet) and the other from the spectator quark. This selection is hereafter referred to as 2J1T. Main backgrounds to this final state are $\mytt$, $\Wb$ boson produced in association with jets ($\Wb$+jets) and QCD multijet events. 

\begin{figure}[!h]
 \centering
	\includegraphics[width=0.35\textwidth]{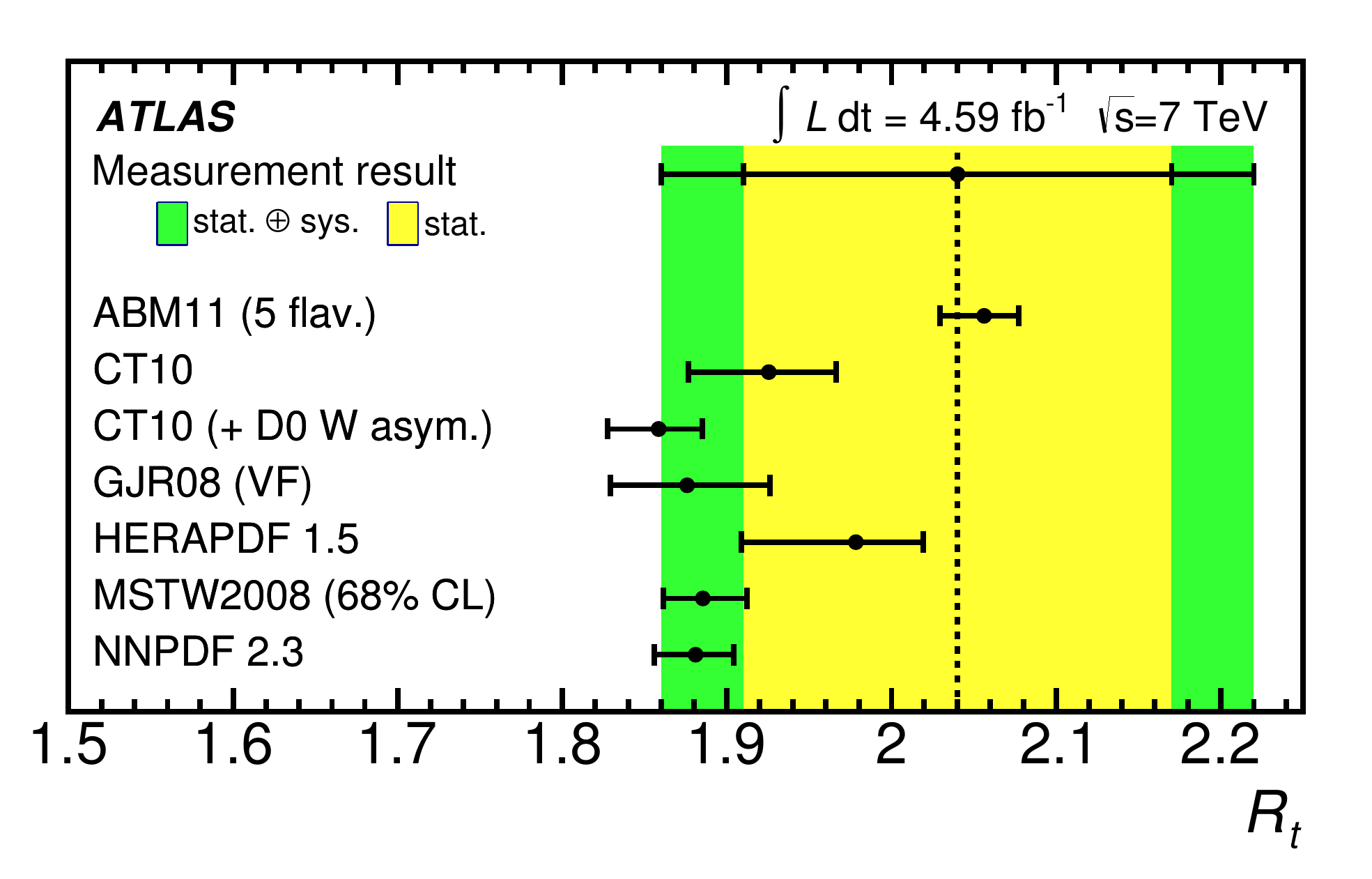} 
	\includegraphics[width=0.35\textwidth]{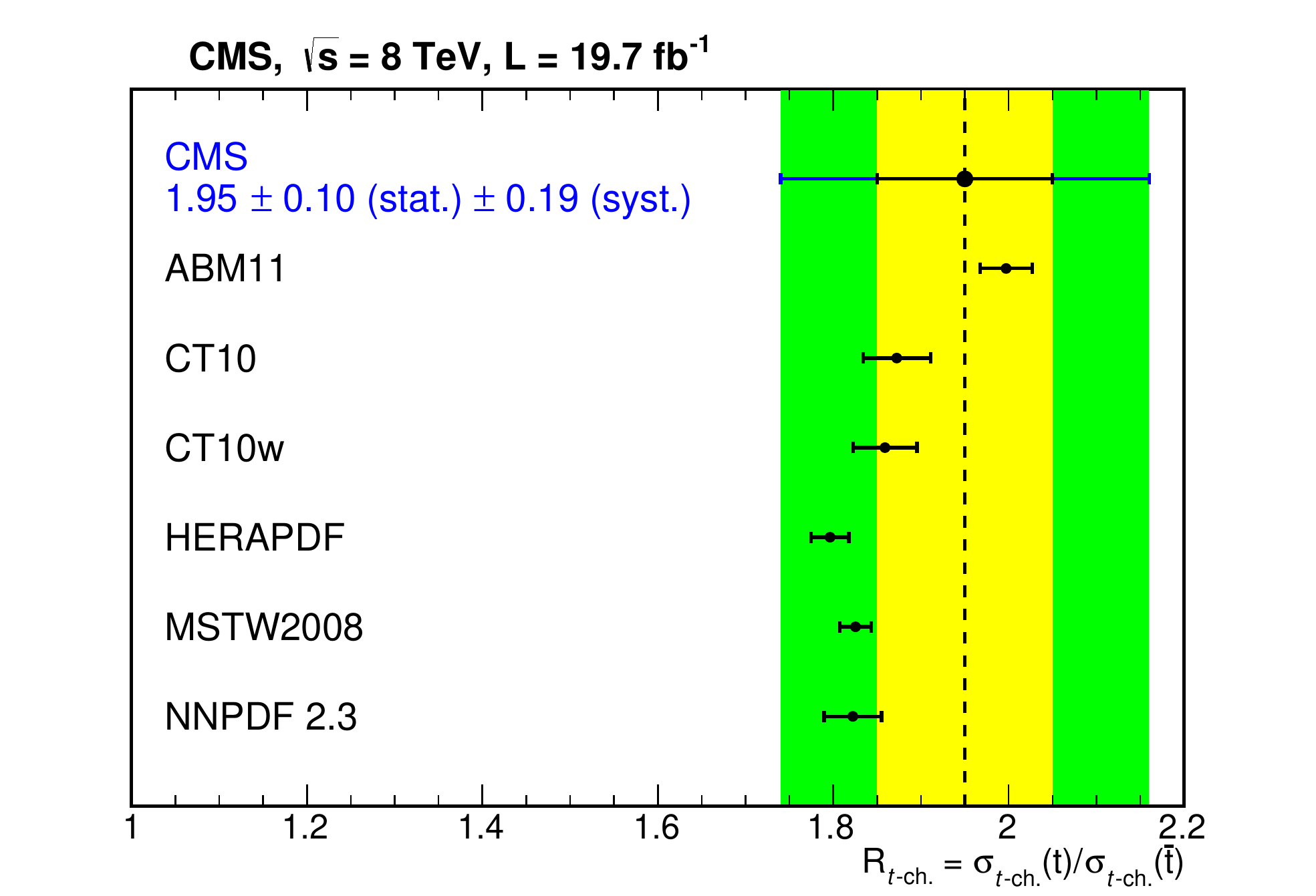} 
    \includegraphics[width=0.35\textwidth]{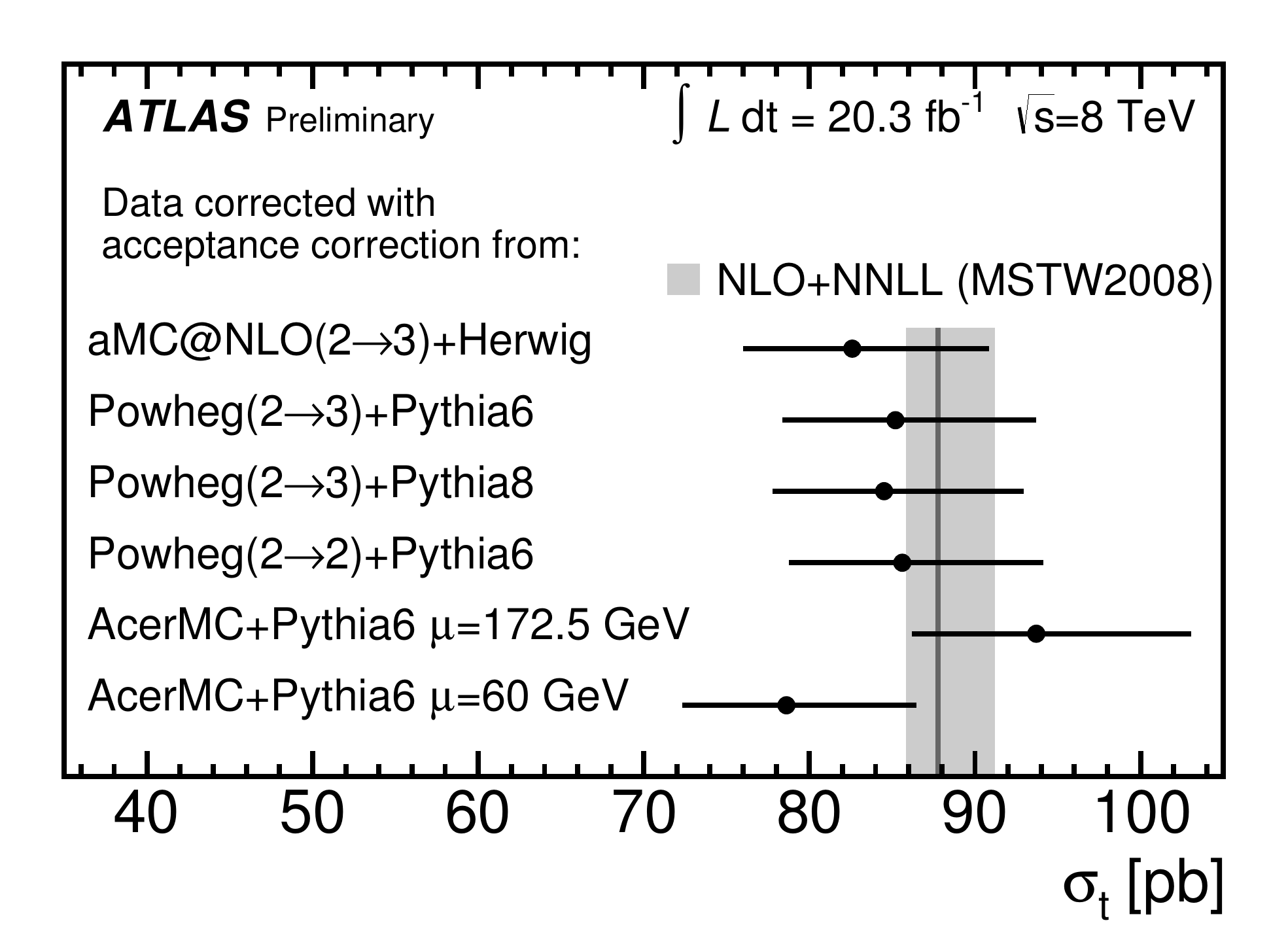}
  \caption{The top-anti-top cross section ratio measurement by ATLAS (top left) at 7\,$\TeV$~\cite{ATLAStchan7} and by CMS (top right) at 8\,$\TeV$~\cite{CMStchan8} together with the extrapolated ATLAS fiducial cross section with different generators at 8\,$\TeV$ (bottom)~\cite{ATLAStchan8}. }
  \label{fig:topmass}
\end{figure}

At $7\,\TeV$ center-of-mass energy, ATLAS has performed a comprehensive analysis~\cite{ATLAStchan7} using $4.6\,\rm fb^{-1}$ of the LHC data. For the signal, the 2J1T and 3J1T event samples are divided into two categories based on the lepton charge. An additional selection on the lepton-jet topology is applied to reject the QCD background. The modelings of simulated backgrounds are validated in control regions with less stringent b tagging criteria. The presence of b jets, as defined in the signal region, is vetoed. The b tagging efficiency is controlled in a 3J2T region. A neural network output is fitted to data simultaneously in the four signal categories to extract the $\mandelt$-channel cross section. The QCD multijet background is estimated using data-driven techniques while other background expectations are derived using simulation. The systematic uncertainties are evaluated with pseudo-experiments. The measurement, with the total uncertainty, yields $\sigma_{\tq q} = 46 \pm 6$\,pb and $\sigma_{\bar{\tq}\rm q} = 23 \pm 4$\,pb for the top quark and antiquark production, respectively, assuming $m_{\rm t} = 172.5\,\GeV$. Accounting for correlations, the top-anti-top cross section ratio is measured to be $R_t = 2.04\pm 0.18$. The jet energy scale (JES) is the dominant systematic uncertainty for the cross sections while for $R_t$, the parton distribution function (PDF) uncertainty is the largest one. Figure~\ref{fig:topmass} shows the $R_t$ measurement compared to predictions using different PDFs. The total cross section, $68\pm 8$\,pb, is obtained by equating the $\tq q$ and $\bar{\tq}\rm q$ signal strengths in the fit. It is used to extract the $|\vtb|$ value, $|\vtb|=1.02\pm 0.07$, assuming $|\vtb|\gg|V_{\rm td}|,|V_{\rm ts}|$. A lower limit of $|\vtb| > 0.88$ is obtained under the assumption of $|\vtb|\leq 1$. These results are comparable with the earlier CMS single top measurement at 7\,TeV~\cite{CMStchan7}.

A fiducial $\mandelt$-channel cross section measurement is provided by the ATLAS collaboration using $20.3\,\invf$ of the $8\,\TeV$ data~\cite{ATLAStchan8}. A binned maximum likelihood fit is performed to the neural network output in the 2J1T region where backgrounds are treated as nuisance parameters. The $\mytt$ and $\Wb$+jets modelings are validated in the 2J2T and 2J0T regions, respectively. The fiducial phase space is defined close to that of the reconstructed and selected data set. The particle-level objects are constructed from stable particles in the final state, with a very similar definition to the reconstructed objects. The fiducial cross section within the detector acceptance is measured to be $\sigma_{\rm fid} = 3.37\pm0.05 \,{\rm(stat.)}\pm 0.47\, {\rm(syst.)} \pm 0.09\, {\rm(lumi.)}$\,pb. Systematic uncertainties are obtained from pseudo-experiments with the dominant contributions from JES and signal generator. The fiducial measurement is extrapolated to the full phase space using different Monte Carlo generators (Fig.~\ref{fig:topmass}). The inclusive cross section is determined to be $\sigma_{\tq} = 82.6 \pm 1.2 \,{\rm(stat.)} \pm 11.4 \,{\rm(syst.)}\pm 3.1 \,{\rm(PDF)} \pm 2.3 \,{\rm(lumi.)}$, using aMCatNLO+{\sc Herwig}. This leads to $|\vtb|=0.97^{+0.09}_{-0.10}$ with a lower bound of $|\vtb|>0.78$ at 95\% confidence level (CL) when $|\vtb|\leq 1$ is assumed.

The CMS experiment has exploited the discriminating feature of the pseudorapidity of the non-b-tagged jet ($\eta_{j'}$) to measure the $\mandelt$-channel cross section at 8\,$\TeV$~\cite{CMStchan8}. Events with one charged lepton (e or $\mu$) in the 2J1T sample are split into two categories based on the lepton charge. Each event category is further divided into a signal region (SR) with a top quark mass of $130<m_{\ell{\bq}\nu}<220\,\GeV$, and two side bands (SB) with $m_{\ell{\bq}\nu}$ out of this range. The QCD multijet events are rejected with a selection on $\ETslash$ and the transverse mass of the $\Wb$ boson ($m_{\rm T}^{\Wb}$) in the electron and muon channels, respectively. Using a data sample equivalent to $19.7\,\invf$, a template fit is performed on the $\eta_{j'}$ distribution in SR with backgrounds treated as constrained nuisance parameters. Data events in SB are used to extract the $\Wb$+jets template. The shape for $\mytt$ is also corrected, using data-driven correction factors from a 3J2T control region. The shape and normalization of QCD multijets are determined with other data driven techniques. Those systematic uncertainties that are not marginalized in the fit, are evaluated with pseudo-experiments. The cross sections for the top quark and top antiquark production are $\sigma_{\rm t}=53.8\pm1.5 \,{\rm(stat)}\pm4.4\,{\rm(syst)}\, \rm pb$ and $\sigma_{\rm \bar{t}}=27.6\pm 1.3 \,{\rm(stat)}\pm3.7\,{\rm (syst)}\,\rm pb$, respectively. The measured inclusive cross section is $\sigma_{\rm tot}=83.6\pm2.3\,{\rm(stat)}\pm7.4 \,{\rm(syst)}\,\rm pb$ with the signal modeling as the dominant systematic uncertainty. The top-anti-top cross section ratio is evaluated, $R_t=1.95\pm0.10\,{\rm(stat.)}\pm0.19\,{\rm(syst.)}$, where PDF as the largest uncertainty. Figure~\ref{fig:topmass} shows the CMS $R_t$ measurement compared with different PDF predictions. The inclusive cross section at 8\,$\TeV$ is found to be larger than the one at 7\,$\TeV$ by $R_{8/7}=1.24\pm0.08\,{\rm (stat)}\pm0.12\,{\rm(syst.)}$. The $|\vtb|$ measurements at 7\,$\TeV$ and 8\,$\TeV$ are combined and result in $|\vl\vtb|=0.998\pm0.038\,{\rm (exp.)}\pm0.016 {\rm (theo.)}$ where $\vl$ accounts for deviations from the SM in the $\twb$ coupling (see Sec.~\ref{moscow}). 
\section{The cross section measurements of $\Wb$-associated production}
The $\tq\Wb$ production has been observed in CMS in the dilepton final state, using $12.2\,\invf$ of the 8\,$\TeV$ data~\cite{CMStW}. A maximum likelihood fit is performed on a BDT output over all lepton flavor combinations (ee, e$\mu$ and $\mu\mu$) and in three signal regions (1J1T, 2J1T and 2J2T). The shapes for signal and backgrounds are taken from simulation. Systematic uncertainties are treated as constrained nuisance parameters in the fit except for the luminosity and theory uncertainties which are unconstrained. The scale uncertainty and the $\mytt$ modeling are the dominant systematic uncertainties. A significance of $6.1\,\sigma$ is observed for the signal, to be compared with the expected values of $5.4\pm1.4\,\sigma$. This corresponds to $\sigma_{\tq\Wb}=23.4\pm5.4\,\rm pb$ and $|\vtb|=1.03\pm0.12\,{\rm (exp.)}\pm0.04\,{\rm (theo.)}$. Constraining $|\vtb|\leq 1$ leads to $|\vtb|>0.78$ at 95\% CL.

A similar measurement is carried out in ATLAS, in the e$\mu$ final state, using $20.3\,\invf$ of the 8\,$\TeV$ data~\cite{ATLAStW}. A fit to the BDT output is simultaneously performed in the 1J1T and 2J$\geq 1$T regions. Templates for signal and backgrounds are taken from simulation and systematic uncertainties evaluated by means of pseudo-experiments with the main contribution coming from the $\tq\Wb$ and $\mytt$ modeling. The measurement has reached to an observed significance of $4.2\,\sigma$ where $4.0\,\sigma$ was expected. This corresponds to $\sigma_{\tq\Wb}=27.2\pm2.8\,{\rm(stat.)}\pm5.4\,{\rm(syst.)}\,\rm pb$ and $|\vl\vtb|=1.10\pm0.12\,{\rm (exp.)}\pm0.03\,{\rm (theo.)}$. Assuming $|\vl| =1$ and $|\vtb|\leq 1$ lead to $|\vtb|>0.72$ at 95\% CL.

The results of the two experiments are combined with the BLUE method~\cite{LHCtW}. The theory modeling systematic uncertainties are considered fully correlated between CMS and ATLAS, while a 30\% correlation is assigned to the luminosity and a 50\% correlation is taken into account for the b tagging uncertainty. The stability of the combination with these choices is verified by examining different correlation values. The combination leads to $\sigma_{\tq\Wb}=25.0\pm1.4\,{\rm(stat.)}\pm4.4\,{\rm(syst.)}\pm0.7\,{\rm(lumi.)}\,\rm pb$ for the $\tq\Wb$ cross section. The combined measurement for the $\vtb$ CKM matrix element is found to be $|\vl\vtb| = 1.06\pm0.11$ where a lower bound of $|\vtb|>0.79$ is obtained at 95\% CL, under the assumption of $|\vl| =1$ and $|\vtb|\leq 1$.
\section{The $\mandels$-channel cross section measurements}
\begin{figure}[h]
\includegraphics[width=20pc]{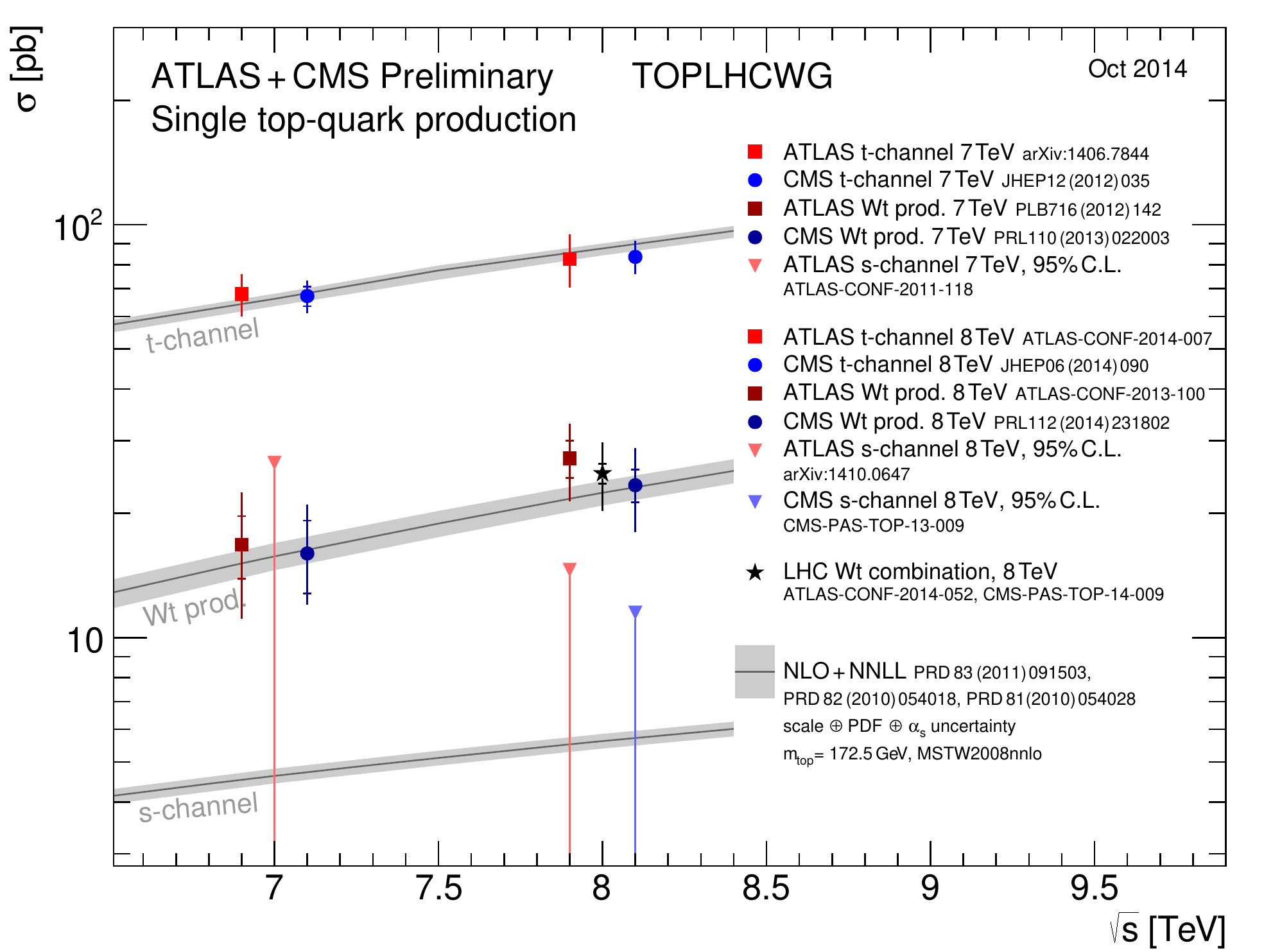}\hspace{2pc}%
\begin{minipage}[b]{15pc}\caption{\label{fig:f4}The LHC summary plot for single-top cross section. The $\mandels$-channel ATLAS result at 8\,$\TeV$ appeared after this presentation, hence not included in the report.}
\end{minipage}
\end{figure}

The $\mandels$-channel signal is characterized by one charged lepton, $\ETslash$ and two b tagged jets. Using $19.3\,\invf$ of the 8\,$\TeV$ data, CMS has set an observed upper limit of $\sigma_{\rm s-ch}<11.5\,\rm pb$ on the $\mandels$-channel cross section~\cite{CMSschan8}. A BDT discriminant is constructed in the 2J2T (signal) and 3J2T ($\mytt$ background) regions. The signal strength is extracted using a likelihood fit to the BDT outputs where the $\mytt$ and $\Wb$+jets backgrounds are constrained in the fit. The QCD multijet background is determined with a similar method to the $\mandelt$-channel analysis. Systematic uncertainties are evaluated using pseudo-experiments. The scale uncertainty is the dominant contribution to the total uncertainty. Figure~\ref{fig:f4} summarizes the LHC single top cross section measurements with 7\,$\TeV$ and 8\,$\TeV$ data samples.
\section{Search for anomalous $\twb$ couplings}
\label{moscow}
Deviations from SM in the $\twb$ vertex can be expressed in terms of the anomalous couplings, $\vl,\vr,\gl$ and $\gr$, presented in this Lagrangian:
\begin{equation}
\label{eq:anomL}
\mathcal{L}_{\twb}^{\rm anom.}=-\frac{g}{\sqrt{2}}\bar{\bq}\gamma^{\mu}(\vl P_{\rm L}+\vr P_{\rm R}){\tq}{\Wb}^{-}_{\mu}-\frac{g}{\sqrt{2}}\bar{\bq}\frac{i\sigma^{\mu\nu}q_{\nu}}{m_{\Wb}}(\gl P_{\rm L}+\gr P_{\rm R}){\tq}{\Wb}^{-}_{\mu}+{\rm H.C.},
\end{equation}
in which $q$ is the difference of the top and bottom quark 4-momenta. A direct search for the anomalous couplings is carried out in CMS using $5.0\,\invf$ of 7\,$\TeV$ data. Signal events with one charged lepton (e or $\mu$) are categorized into sub-samples, namely: 2J1T, 3J1T and 3J2T. QCD multijet events are rejected by imposing a lower bound on the output of a dedicated Bayesian neural network (BNN). The $\mytt$ modeling is validated in a 4J2T sample where 2J0T and 3J0T samples are used to verify the $\Wb$+jets modeling. A "SM BNN" is constructed to discriminate between the SM $\mandelt$-channel and SM backgrounds whereas an "$\rm a\twb$ BNN" is trained for the anomalous hypothetic scenarios against all SM processes. The anomalous scenarios are simulated for ($\vl$,$\gl$) and ($\vl,\vr$) combinations where those couplings that are not present in the combination are set to zero. For each combination the two BNN discriminants are used as inputs in the statistical analysis. The observed (expected) limits at 95\% CL are $|\vl|>0.92(0.88)$ and $|\gl|<0.09(0.06)$ for ($\vl$,$\gl$), and $|\vl|>0.90(0.88)$ and $|\vr|<0.34(0.39)$ for ($\vl$,$\vr$). 
\section{Top quark properties in production and decay}
In the SM, the top quark is highly polarized in the direction of the spectator quark in the $\mandelt$-channel process. Its spin is also correlated with the angular properties of the decay products. The angle between the charged lepton and non-b-tagged jet 3-momenta in the top quark rest frame ($\theta^*$) is used in CMS to measure the top quark polarization~\cite{CMStopPol}. The data sample corresponds to $20\,\invf$ and the event selection is similar to the CMS $\mandelt$-channel cross section measurement at 8\,$\TeV$ with an additional threshold on a boosted decision tree (BDT) output to purify the signal sample. The background-subtracted distribution of $\cos\theta^*$ is then unfolded to particle level. A fit to the unfolded distribution leads to a spin asymmetry of $A_l = 0.42\pm0.07\,{\rm(stat.)}\pm0.15\,{\rm(syst.)}$ for the muon channel and $A_l = 0.31\pm0.11\,{\rm(stat.)}\pm0.23\,{\rm(syst.)}$ for the electron channel. The top quark polarization, $A_l = 0.82\pm0.12\,{\rm(stat.)}\pm0.32\,{\rm(syst.)}$, is extracted from the combination of the two measurements using the BLUE method.

The $\Wb$ helicity fractions are sensitive to the anomalous couplings in Eq.~\ref{eq:anomL}. The helicity angle $\theta^{*}_{\ell}$ is defined as the angle between the $\Wb$ boson momentum in the top quark rest frame and the momentum of the down-type decay fermion in the rest frame of the $\Wb$ boson. The functional form of the top quark partial decay width can be written as $\rho(\cosTheta|\vec{F})$, representing the contributions from the right-handed ($\fr$), left-handed ($\myfl$) and longitudinal (\fz) helicity fractions of the $\Wb$ boson. A reweighting method, built on $\rho(\cosTheta|\vec{F})$, is employed to measure the $\Wb$ helicity fractions~\cite{CMSWhel}, with the same data sample and event selection as in Ref.~\cite{CMStchan8}. All events containing a $\tq\to\ell\bq\nu$ interaction, including $\mytt$ and other single top processes, are considered as signal, reweighted and added to backgrounds. The resulting $\cosTheta$ distribution is fitted to the data to extract, simultaneously, the $\Wb$ helicity fractions and the $\Wb$+jets background contamination. The combination of the muon and electron channels yields $\cresfl$, $\cresfz$ and $\cresfr$, all consistent with the SM predictions. The measured fractions are used to set limits on the $\gl$ and $\gr$ anomalous couplings.	

\section*{Acknowledgement}
The author would like to thank the ATLAS and CMS collaboration for their incredible work in single top quark studies. Also thanks to FNRS (Fond National de la Recherche Scientifique) from Belgium who financially supported the author for this conference.


\section*{References}
\begin{thebibliography}{9}
\bibitem{JINSTLHC} L. Evans and P. Bryant, “LHC Machine”, \textit{JINST} \textbf{3} (2008) S08001, doi:10.1088/1748-0221/3/08/S08001.
\bibitem{JINSTCMS} CMS Collaboration, “The CMS experiment at the CERN LHC”, \textit{JINST} \textbf{3} (2008) S08004, doi:10.1088/1748-0221/3/08/S08004.
\bibitem{JINSTATLAS} ATLAS Collaboration, “The ATLAS Experiment at the CERN Large Hadron Collider”, \textit{JINST} \textbf{3} (2008) S08003, doi: 10.1088/1748-0221/3/08/S08003.
\bibitem{ATLAStchan7} ATLAS Collaboration, “Comprehensive measurements of $\mandelt$-channel single top-quark production cross sections at $\sqrt{s}$=7\,$\TeV$ with the ATLAS detector”, (2014), arXiv:1406.7844.
\bibitem{CMStchan7} CMS Collaboration, “Measurement of the $\mandelt$-channel single top quark production cross section in pp collisions at $\sqrt{s}$=7\,$\TeV$”, \textit{JHEP} \textbf{12} (2012) 035, doi:10.1007/JHEP12(2012)035, arXiv:1209.4533.
\bibitem{ATLAStchan8} ATLAS Collaboration, “Measurement of the Inclusive and Fiducial Cross-Section of Single Top-Quark $\mandelt$-Channel Events in $pp$ Collisions at $\sqrt{s}$=8\,$\TeV$”, (2014), ATLAS-CONF-2014-007.
\bibitem{CMStchan8} CMS Collaboration, “Measurement of the $\mandelt$-channel single-top-quark production cross section and of the $\vtb$ CKM matrix element in pp collisions at $\sqrt{s}$=8\,$\TeV$”, \textit{JHEP} \textbf{06} (2014) 090, doi:10.1007/JHEP06(2014)090.
\bibitem{CMStopPol} CMS Collaboration, “Measurement of top-quark polarization in $\mandelt$-channel single-top production”, (2013), CMS-PAS-TOP-13-001.
\bibitem{CMSWhel} CMS Collaboration, “Measurement of the $\Wb$ boson helicity in events with a single reconstructed top quark in pp collisions at $\sqrt{s}$=8\,$\TeV$”, (2014), arXiv:1410.1154.
\bibitem{CMStW} CMS Collaboration, “Observation of the Associated Production of a Single Top Quark and a $\Wb$ Boson in pp Collisions at $\sqrt{s}$=8\,$\TeV$”, \textit{Phys. Rev. Lett.} \textbf{112} (2014) 231802, doi:10.1103/PhysRevLett.112.231802, arXiv:1401.2942.
\bibitem{ATLAStW} ATLAS Collaborations, “Measurement of the cross-section for associated production of a top quark and a $\Wb$ boson at $\sqrt{s}$=8\,$\TeV$ with the ATLAS detector”, (2013), ATLAS-CONF-2013-100.
\bibitem{LHCtW} ATLAS and CMS Collaborations, “Combination of cross-section measurements of associated production of a single top-quark and a $\Wb$ boson at $\sqrt{s}$=8\,$\TeV$ with the ATLAS and CMS experiments”, (2014), ATLAS-CONF-2014-052, CMS-PAS-TOP-14-009.
\bibitem{CMSschan8} CMS Collaboration, “Search for $\mandels$-channel single top-quark production in pp collisions at $\sqrt{s}$=8\,$\TeV$”, (2013), CMS-PAS-TOP-13-009.
\end{thebibliography}
\end{document}